\documentclass[aps,prl,showpacs,twocolumn,superscriptaddress,letterpaper]{revtex4}
\usepackage{amsmath}
\usepackage{amssymb}
\usepackage{amsfonts}
\usepackage{graphicx}% Include figure files

\begin{document}
\title{Pairing Symmetry in Layered BiS$_2$ Compounds Driven by Electron-Electron Correlation}

\author{Yi Liang}
\affiliation{Institute of Physics, Chinese Academy of Sciences, Beijing 100190, China}
\author{Xianxin Wu}
\affiliation{Institute of Physics, Chinese Academy of Sciences, Beijing 100190, China}

\author{Wei-Feng Tsai}
\affiliation{Department of Physics, National Sun Yat-sen University, Kaohsiung 804, Taiwan}

\author{Jiangping Hu}
\affiliation{Institute of Physics, Chinese Academy of Sciences, Beijing 100190,
China}
\affiliation{Department of Physics, Purdue University, West Lafayette, Indiana 47907, USA}

%\date{\today}

\begin{abstract}
We investigate the pairing symmetry of layered BiS$_2$ compounds by assuming that electron-electron correlation is still important so that the pairing is rather short range. We find that the extended $s$-wave pairing symmetry always wins over $d$-wave when the pairing is confined between  two short range sites up to next nearest neighbors.  The pairing strength is peaked around the doping level $x=0.5$, which is consistent with experimental observation. The  extended $s$-wave pairing symmetry is very robust against spin-orbital coupling because it is mainly determined by the structure of Fermi surfaces.  Moreover, the extended $s$-wave pairing can be distinguished from conventional $s$-wave pairing by measuring and comparing superconducting gaps of different Fermi surfaces.

\end{abstract}


\maketitle

\section*{Introduction}
Very recently, Mizuguchi \emph{et al.} discovered a new layered superconductor Bi$_4$O$_4$S$_3$ with $T_{c}\sim 4.5$ K\cite{Mizuguchi1} and other groups also reported similar materials,
LaO$_{1-x}$F$_{x}$BiS$_2$\cite{Mizuguchi2} and NdO$_{1-x}$F$_{x}$BiS$_2$\cite{Demura, Jha}. These materials share some common features with both copper-based and iron-based high temperature superconductors. The electronic properties of these materials are determined by the BiS$_2$ layer\cite{Mizuguchi1}, similar to the CuO$_2$ layer in cuprates or the
Fe$_2$An$_2$(An=P, As, Se, Te) layer in iron-based superconductors.  Electron doping to the BiS$_2$ layer can induce superconductivity.  For example,  in
LaO$_{1-x}$F$_{x}$BiS$_2$,  the partial replacement of O by F provides
electron doping and induces superconductivity.
%It  has been shown that superconductivity is developed when the doping level is  higher than 0.4e per Bi atom in LaO$_{1-x}$F$_{x}$BiS$_2$ and 0.1e per Bi atom in NdO$_{1-x}$F$_{x}$BiS$_2$.
Some properties in the superconducting (SC) state are also shown to be unconventional although it is too early to make a conclusive statement\cite{Li}.   A number of studies appear shortly after the discovery of BiS$_2$ layered superconductor\cite{Tan,Kotegawa,Usui}.

The basic band structure of these materials has been calculated by first principle calculation. The conduction band in the BiS$_2$ layer is mainly attributed to the $p_x$ and $p_y$ orbits
of Bi\cite{Usui}. Neglecting the interlayer coupling and states away from Fermi surface, Usui \emph{et al.}\cite{Usui} proposed a two-orbital model, in which only $p_x$ and $p_y$ orbitals are included, to describe the band structure. A good Fermi-surface nesting and
two quasi-one-dimensional bands have been found in
LaO$_{1-x}$F$_{x}$BiS$_2$\cite{Usui}. The importance of Fermi surface nesting has been emphasized in Ref.~\onlinecite{Kotegawa2}.

Because of the low $T_c$ and the extended $p$-orbitals in BiS$_2$ layered materials, electron-phonon coupling was suggested to play the main role in the Cooper pairing\cite{Usui,Wan}. However, because electron-electron correlation generally is more important in a low dimensional system, the correlation effect might play an important role in driving superconductivity even if the $p$-orbitals of Bi are much less localized compared with $d$-orbitals in cuprates and iron-based superconductors\cite{Usui,Zhou}. Experimentally it has also been proposed in Ref.~\onlinecite{Li} that the SC pairing is strong and exceeds the limit of the phonon mediated picture. The correlation effect, therefore, seems to be a good candidate responsible for the SC pairing in these materials.

In the two-orbital model based on the $p$-orbitals of Bi, the quasi-one dimensional bands  are generated through the $p$-orbitals of S\cite{Usui}. If electron-electron correlation is important, the superexchange mechanism can naturally result in a next-nearest-neighbor (NNN) antiferromagnetic (AFM)  exchange coupling.  Although the existence of AFM  fluctuation is still needed to be justified experimentally, theoretically it is legitimate to search for the consequence of possible correlation effect in these materials.

In this paper, we investigate the pairing symmetry of layered BiS$_2$ compounds by assuming that electron-electron correlation is still important so that the pairing is rather short range. Under the assumption  that the short range pairing stems from short AFM exchange couplings, we find that the extended $s$-wave pairing symmetry always wins over $d$-wave. Such result is very similar to the case in iron-based superconductors\cite{Hosono, hirschfeld, seo2008}. We find that the pairing strength is peaked around the doping level $x=0.5$, which is also consistent with experimental observation. The pairing symmetry is very robust against spin-orbital coupling because it is mainly determined by the structure of Fermi surfaces. The extended $s$-wave pairing can be tested by measuring the gap distributions at different Fermi surfaces.

\section*{Theoretical Model}
LaOBiS$_{2}$ is an indirect semiconductor. Its band gap is about 0.82 eV, which is much larger than the superconducting gap, $T_{c}\sim 10$ K. The simplest effective model is a two-orbital model constructed by  $6p_{x}$ and $6p_{y}$ orbitals of Bi. As shown in band structure of LaO$_{0.5}$F$_{0.5}$BiS$_{2}$ from first principle calculation\cite{Usui}, the four lowest conduction bands are splitted slightly. The splitting is so small compared with their band width so that the low energy physics can be captured by a two dimensional two-orbital model that ignores the coupling between BiS$_{2}$ layers. We adopt the band structure proposed in Ref.~\onlinecite{Usui}, in which the noninteracting Hamiltonian reads
$H_{0}=\sum_{\mathbf{k},\sigma}\Psi_{\mathbf{k},\sigma}^{\dagger}T(\mathbf{k})\Psi_{\mathbf{k},\sigma}$,
\begin{equation}
T(\mathbf{k})=\left(\begin{array}{cc}
\epsilon_{X}(\mathbf{k})-\mu & \epsilon_{XY}(\mathbf{k})\\
\epsilon_{XY}(\mathbf{k})^{*} & \epsilon_{Y}(\mathbf{k})-\mu\end{array}\right),
\end{equation}
where $\Psi_{\mathbf{k},\sigma}^{\dagger}=(C_{\mathbf{k},\sigma,X,}^{\dagger},C_{\mathbf{k},\sigma,Y}^{\dagger})$
are creation operators for electrons with spin $\sigma$
in the two orbitals $p_{X},p_{Y}$ aligned along {\it diagonal} directions in Bi unit cell and
\begin{eqnarray}
\epsilon_{X}(\mathbf{k}) & = & 2t(\cos(k_{x})+\cos(k_{y}))+2t^\prime \cos(k_{x}+k_{y})\nonumber \\
 &  & +2t^{\prime\prime}\cos(k_{x}-k_{y})\nonumber \\
 &  & +2t_{21}(\cos(2k_{x}+k_{y})+\cos(k_{x}+2k_{y}))\nonumber \\
 &  & +2t_{21}^{\prime}(\cos(2k_{x}-k_{y})+\cos(-k_{x}+2k_{y})),\nonumber \\
\epsilon_{Y}(\mathbf{k}) & = & 2t(\cos(k_{x})+\cos(k_{y}))+2t^{\prime}\cos(k_{x}-k_{y})\nonumber \\
 &  & +2t^{\prime\prime}\cos(k_{x}+k_{y})\nonumber \\
 &  & +2t_{21}(\cos(2k_{x}-k_{y})+\cos(-k_{x}+2k_{y}))\nonumber \\
 &  &+2t_{21}^{\prime}(\cos(2k_{x}+k_{y})+\cos(k_{x}+2k_{y})),\nonumber \\
\epsilon_{XY}(\mathbf{k}) & = & 2t_{XY}(\cos(k_{x})-\cos(k_{y}))\nonumber \\
&  & +2t_{20XY}(\cos(2k_{x})-\cos(2k_{y}))\nonumber \\
 &  & +2t_{21XY}(\cos(2k_{x}+k_{y})-\cos(k_{x}+2k_{y})\nonumber \\
 &  &  +\cos(2k_{x}-k_{y})-\cos(k_x-2k_y)).
\label{eq2}\end{eqnarray}
The hopping parameters  in  Eq.~\ref{eq2}  are given by $t=-0.167$, $t^{\prime}=0.88$, $t^{\prime\prime}=-0.094$, $t_{XY}=0.107$, $t_{21}=0.069$, $t_{21}^{\prime}=0.014$, $t_{20XY}=-0.028$, and $t_{21XY}=0.02$ in units of eV.
It is clear from these parameters that the material is quasi-one dimensional because the value of $t^\prime$ is almost an order of magnitude larger than other hopping parameters.  The model has a Lifshitz transition at electron doping $x=0.452$ and $x=0.515$ where the Fermi surface topologies are changed.

\begin{figure}%[t]
 \includegraphics[height=4.0cm]{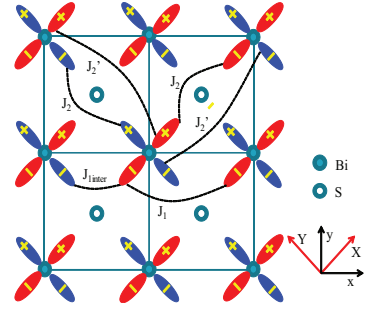}
 \caption{The orbital basis and interaction in our calculation}\label{basis}
 \end{figure}

If the electron-electron correlation effect can not be ignored, the hopping through the $p$-orbitals of S would naturally lead to an AFM exchange coupling between two NNN sites of Bi for the above two-orbital models. Taking the standard approach, in general, we can write the interacting Hamiltonian as
\begin{equation}
H_{I}=\sum_{\langle\langle i,j\rangle\rangle,\alpha}J_{ij,\alpha}(\mathbf{S}_{i,\alpha}\cdot\mathbf{S}_{j,\alpha}-
\frac{1}{4}n_{i,\alpha}n_{j,\alpha}),
\end{equation}
where $\mathbf{S}_{i,\alpha}=\frac{1}{2}\sum_{\sigma\sigma^\prime}C_{i,\sigma,\alpha}^{\dagger}\vec{\sigma}_{\sigma\sigma^\prime}
C_{i,\sigma^\prime,\alpha}$ is the local spin operator, $n_{i,\alpha}$ is the local density operator, $\langle\langle i,j\rangle\rangle$ denotes a pair of NNN sites, $\alpha$ is orbital index and we use $X,Y$ to label $p_{X}$ and $p_{Y}$ orbitals, respectively. Owing to the quasi-one dimensional property, the AFM NNN coupling can take two independent values such that $J_{2,X}=J_{2,Y}^\prime=J_{2}$ and $J_{2,X}^\prime=J_{2,Y}=J_{2}^\prime$, keeping C$_4$ symmetry unbroken, with $J_2$ expected to dominate over other exchange parameters.
If we rewrite the interaction term in momentum space,
\begin{eqnarray}
&& H_{I}=\sum_{\mathbf{k},\mathbf{k}^\prime,\alpha}V_{\mathbf{k},\mathbf{k}^\prime}^{\alpha}
C_{\mathbf{k},\uparrow,\alpha}^{\dagger}C_{\mathbf{-k},\downarrow,\alpha}^{\dagger}
C_{-\mathbf{k}^\prime,\downarrow,\alpha}C_{\mathbf{k}^\prime,\uparrow,\alpha}, \nonumber \\
&& V_{\mathbf{k},\mathbf{k}^\prime}^{\alpha}=\frac{-2}{N}(J_{2,\alpha}\cos(k_{x}+k_{y})
\cos(k_{x}^\prime+k_{y}^\prime)\nonumber \\
&&
+J_{2,\alpha}^\prime\cos(k_{x}-k_{y})\cos(k_{x}^\prime-k_{y}^\prime)).
\end{eqnarray}

In general, the presence of electron-electron correlation can also significantly renormalize the bare band structure. In the strong coupling limit, the band width renormalization strongly depends on doping. However, here since  the correlation is, at most, moderate, we assume that the renormalization does not vary significantly as the function of doping. Under such an assumption, we can use standard  mean-field approximation to obtain the SC state and its favored pairing symmetry. This approach has been used to study iron-based superconductors where the correlation  is most likely around intermediate coupling strength. Under mean-field approximation, the total Hamiltonian can be simplified as
\begin{eqnarray}
& &  H^{MF}=\sum_{\mathbf{k}}\phi_{\mathbf{k}}^{\dagger}A(\mathbf{k})\phi_{\mathbf{k}}+\sum_{\mathbf{k}}\left(\epsilon_{X}(\mathbf{-k})+\epsilon_{Y}(-\mathbf{k})-2\mu\right)\nonumber \\
&& +\frac{N}{2J_{2}}(\Delta_{X}^{*}\Delta_{X}+\Delta_{Y}^{*}\Delta_{Y}+\Delta_{X}^{'*}\Delta_{X}+\Delta_{Y}^{'*}\Delta_{Y}),
\end{eqnarray}
where
\begin{equation}A(\mathbf{k})  =  \left(\begin{array}{cccc}
\tilde{\epsilon}_{X}(\mathbf{k}) & \epsilon_{XY}(\mathbf{k}) & \Delta_{XX}(\mathbf{k})^{*} & 0\\
\epsilon_{XY}(\mathbf{k})^{*} & \tilde{\epsilon}_{Y}(\mathbf{k}) & 0 & \Delta_{YY}(\mathbf{k})^{*}\\
\Delta_{XX}(\mathbf{k}) & 0 & -\tilde{\epsilon}_{X}(\mathbf{k}) & -\epsilon_{XY}(\mathbf{k})^{*}\\
0 & \Delta_{YY}(\mathbf{k}) & -\epsilon_{XY}(\mathbf{k}) & \tilde{\epsilon}_{Y}(\mathbf{k})\end{array}\right)
\end{equation}
with $\phi_{\mathbf{k}}^{\dagger}=(C_{\mathbf{k},\uparrow,X}^{\dagger},C_{\mathbf{k},
\uparrow,Y}^{\dagger},C_{\mathbf{-k},\downarrow,X},C_{\mathbf{-k},\downarrow,Y})$
, $\tilde{\epsilon}_{\alpha=X,Y}=\epsilon_{\alpha}-\mu $, $d_{\mathbf{k}^\prime,\uparrow,\alpha}=\left\langle C_{-\mathbf{k}^\prime,\downarrow,\alpha}C_{\mathbf{k}^\prime,\uparrow,\alpha}\right\rangle $ and
\begin{eqnarray}
\Delta_{\alpha\alpha}(\mathbf{k}) & = & \Delta_{\alpha}\cos(k_{x}+k_{y})+\Delta_{\alpha}^\prime\cos(k_{x}-k_{y}),\\
\Delta_{\alpha} & = & \left[\frac{-2}{N}J_{2,\alpha}\sum_{\mathbf{k}^\prime}d_{\mathbf{k}^\prime,\uparrow,\alpha}
\cos(k_{x}^\prime+k_{y}^\prime)\right],\\
\Delta_{\alpha}^\prime & = & \left[\frac{-2}{N}J_{2,\alpha}^\prime\sum_{\mathbf{k}^\prime}
d_{\mathbf{k}^\prime,\uparrow,\alpha}\cos(k_{x}^\prime-k_{y}^\prime)\right].
\end{eqnarray}

Since we only consider spin singlet pairing, there are two possible pairing symmetries. In BiS2 layered  materials ( two dimensional system ), C4 rotation symmetry along z-axis is respected. S-wave pairing is defined if  the phase of the order parameter  has no change under C4 rotation and if  the phase changes, then the system is in d-wave pairing. In real space,  which can be obtained from Eq.8,9 by Fourier transformation. Under C4 rotation, $\Delta_{X}$ is changed into $\Delta_{Y}^\prime$ and $\Delta_{Y}$ into $\Delta_{X}^\prime$. Thus, in our definition, s-wave pairing is taken if $\Delta_{X}=\Delta_{Y}^\prime$ and $\Delta_{X}^\prime=\Delta_{Y}$ while the $d$-wave pairing is chosen if $\Delta_{X}=-\Delta_{Y}^\prime,\Delta_{X}^\prime=-\Delta_{Y}$. The above equations can be solved self-consistently with standard approach. By diagonalizing $A(\mathbf{k})$ via an unitary transformation, $U^\dagger(\mathbf{k})A(\mathbf{k})U(\mathbf{k})$, we obtain four Bogoliubov quasi-particle eigenvalues $E_{1}=-E_{3}$ and $E_{2}=-E_{4}$, which are given by
\begin{widetext}
%\begin{eqnarray}
%E_{m=1,2}(\mathbf{k}) & = & \frac{1}{\sqrt{2}}\sqrt{(\tilde{\epsilon}_{X}^{2}+\tilde{\epsilon}_{Y}^{2}+2|\epsilon_{XY}|^{2}+\Delta_{XX}^{2}+\Delta_{YY}^{2})\pm}\nonumber \\
% &  & \overline{\pm\sqrt{(\tilde{\epsilon}_{X}^{2}-\tilde{\epsilon}_{Y}^{2}+\Delta_{XX}^{2}-\Delta_{YY}^{2})^{2}+4|\epsilon_{XY}|^{2}[(\tilde{\epsilon}_{X}+\tilde{\epsilon}_{Y})^{2}+\Delta_{XX}^{2}+\Delta_{YY}^{2}]-4\Delta_{XX}\Delta_{YY}(\epsilon_{XY}^{2}+\epsilon_{XY}^{*2})}}.\nonumber \\
% &  & \hspace{1em}\end{eqnarray}
\begin{eqnarray}
E_{m=1,2}(\mathbf{k}) & = & \frac{1}{\sqrt{2}}\Bigl\{(\tilde{\epsilon}_{X}^{2}+\tilde{\epsilon}_{Y}^{2}+2|\epsilon_{XY}|^{2}+\Delta_{XX}^{2}+\Delta_{YY}^{2})\pm \nonumber \\
 &  & \pm\sqrt{(\tilde{\epsilon}_{X}^{2}-\tilde{\epsilon}_{Y}^{2}+\Delta_{XX}^{2}-\Delta_{YY}^{2})^{2}+4|\epsilon_{XY}|^{2}[(\tilde{\epsilon}_{X}+\tilde{\epsilon}_{Y})^{2}+\Delta_{XX}^{2}+\Delta_{YY}^{2}]-4\Delta_{XX}\Delta_{YY}(\epsilon_{XY}^{2}+\epsilon_{XY}^{*2})}\Bigr\}^\frac{1}{2}.\nonumber \\
 &  & \hspace{1em}\end{eqnarray}
 \end{widetext}
The self-consistent gap equations are
\begin{eqnarray}
\Delta_{X(Y)} & = & \sum_{\mathbf{k},m}\frac{-2}{N}J_{2,X(Y)}\cos(k_{x}+k_{y})\nonumber\\
&  & \hspace{1em} U_{3(4),m}^{*}(\mathbf{k})U_{1(2),m}(\mathbf{k})F[E_{m}(\mathbf{k})]\\
\Delta_{X(Y)}^\prime & = & \sum_{\mathbf{k},m}\frac{-2}{N}J_{2,X(Y)}^\prime\cos(k_{x}-k_{y})\nonumber\\
&  & \hspace{1em} U_{3(4),m}^{*}(\mathbf{k})U_{1(2),m}(\mathbf{k})F[E_{m}(\mathbf{k})]\end{eqnarray}
where $F[E]$ is Fermi-Dirac distribution function, $F[E]=1/(e^{E/k_{B}T}+1)$.

\section*{Results and discussion}
\begin{figure}%[t]
 \includegraphics[height=4.0cm]{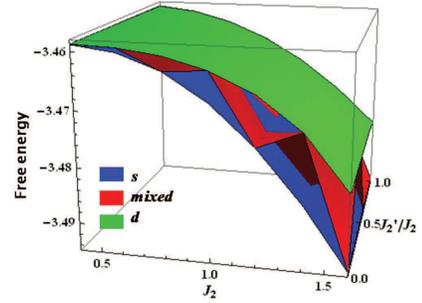}
 \caption{The free energy of $s$- and $d$-wave pairing states as a function of $J_{2},J_{2}^\prime$ at $0.51$ electron doping.}\label{freeEnergy}
 \end{figure}

\begin{figure}%[b]
 \includegraphics[height=5.2cm]{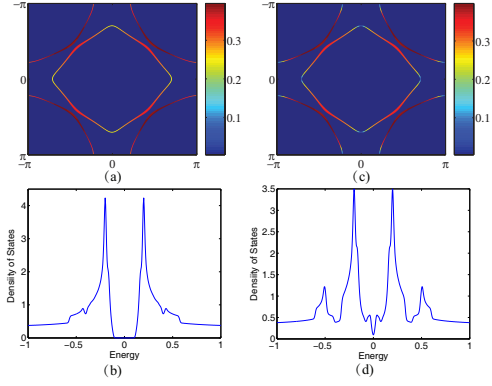}
 \caption{The SC gap distribution over the whole Fermi surface with $\Delta_X=0.2$ at $0.51$ electron doping: (a) for $s$-wave pairing state and (c) for $d$-wave pairing state. The line width of Fermi surface is proportional to  $[\frac{d\epsilon(\mathbf{k})}{dk_{\bot}}]^{-1}$, where $\epsilon(\mathbf{k})$ is the eigenvalue of %$H_0$
$T(\mathbf{k})$ and $\frac{d\epsilon(\mathbf{k})}{dk_{\bot}}$ represents the derivative of $\epsilon(\mathbf{k})$ along normal direction of Fermi surface. The density of states (DOS) in SC ground state: (b) for $s$-wave pairing state and (d) for $d$-wave pairing state. The finite DOS at Fermi level in (d) is just numerical error, induced by the large smearing parameter.}\label{overlap051}
 \end{figure}

The above equations can be solved numerically. In Fig.~\ref{freeEnergy}, we plot the  free energy of $s$- and $d$-wave pairing states in $J_{2}-J_{2}^\prime/J_2$ plane with $0.51$ electron doping. In all plotting region, the extended $s$-wave is favored. The reason that the extended $s$-wave is favored over the $d$-wave can be roughly understood by the overlap between the absolute value of the  pairing gap form factor with the Fermi surface topologies as shown in Ref.~\onlinecite{Hu}. For simplicity, we consider the case that only $J_2$ exists. We first assume the $s$-wave and $d$-wave have the same pairing amplitude, then the overlap between SC gap and Fermi surface is defined as the sum of the absolute values of the SC gap form factor over the whole Fermi surface. The pairing channel is the one that has largest overlap. Fig.~\ref{overlap051} illustrates the SC gap distribution over the entire Fermi surface with $\Delta_X=0.2$ at doping $x=0.51$. The gap distributions of $s$- and $d$-wave pairing states are the same except the region near the nodes of $d$-wave state. It's obvious that the overlap of $s$-wave pairing state is  larger than that of $d$-wave pairing state. Our calculation also shows that the $s$-wave pairing does not co-exist with $d$-wave pairing.  As shown in Fig.~\ref{freeEnergy}, the mixed $s$-wave and $d$-wave state has higher free energy than the pure $s$-wave state.

 \begin{figure}%[t]
 \includegraphics[height=7.0cm]{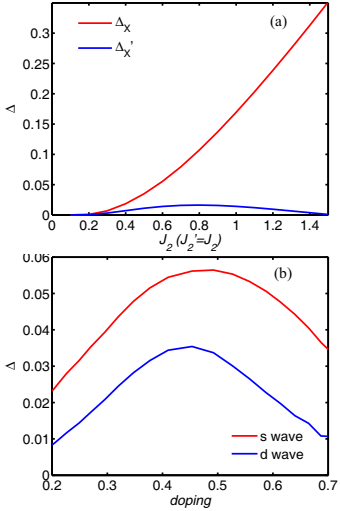}
 \caption{(a) The amplitude of pairing gap induced by $J_{2}$, $\Delta_X$, and that by $J_{2}^\prime$, $\Delta_X^\prime$, as a function of $J=J_{2}=J_{2}^\prime$ at $0.51$ electron doping. (b) $\Delta_X$ as a function of doping at $J_{2}=0.6,J_{2}^\prime=0$}\label{DX_muJ2}
 \end{figure}

The extended $s$-wave pairing is very robust against the variation of parameters. This is because the quasi-one dimensional nature in the band structure results in the dominant pairing channel which is mainly from the $J_2$ term. For example, even we increase $J_2^\prime$, the amplitude of pairing induced by $J_{2}$ is still much larger than that by $J_{2}^\prime$, as shown Fig.~\ref{DX_muJ2} (a). $\Delta_X$ is five times of $\Delta_X^\prime$
when  $J_2=0.6, J_{2}^\prime=J_2$ and ten times when $J_2=1, J_{2}^\prime=J_2$. Furthermore, the optimal doping to achieve the highest $T_c$ is around 0.5. Therefore, given $J_2$, we will obtain a superconducting dome with optimal $T_c$ at $x\sim 0.5$ as shown in Fig.~\ref{DX_muJ2} (b). These results are consistent with the experimental results.  Experimentally, the optimal doping of LaO$_{1-x}$F$_x$BiS$_2$ is suggested to be $\sim0.5$ per Bi atom\cite{Mizuguchi2} and $\sim0.3$ per Bi atom for NdO$_{1-x}$F$_x$BiS$_2$\cite{Demura}. In the later case,   the actual value may be higher because of the existence of impurity phases\cite{Mizuguchi2, Jha}.

\begin{figure}%[t]
 \includegraphics[height=4.0cm]{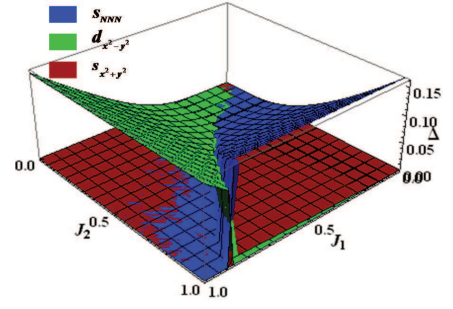}
 \caption{The amplitudes of different pairing gap components induced by $J_1,J_{2}$ with $J_2'=0$ at electron doping$=0.51$. $s_{NNN}$ is the s wave resulting from NNN AFM coupling, $s_{x^2+y^2}$ and $d_{x^2-y^2}$ from NN AFM coupling. }\label{J1J2}
 \end{figure}

\begin{figure}%[t]
 \includegraphics[height=4.0cm]{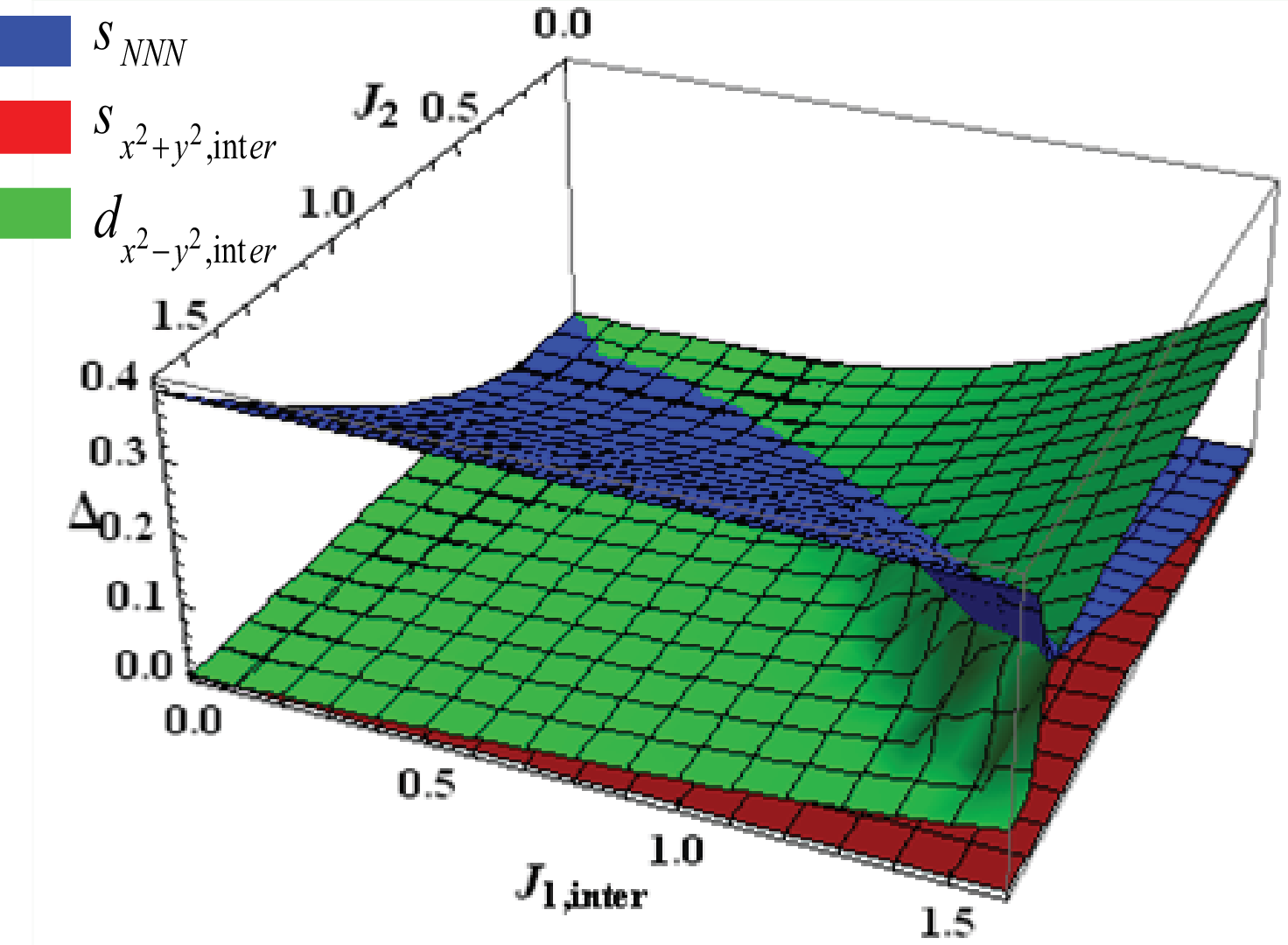}
 \caption{The amplitudes of different pairing gap components induced by $J_{1,inter},J_{2}$ with $J_1=0, J_2'=0$ at electron doping$=0.51$. $s_{NNN}$ is the s wave resulting from NNN AFM coupling, $s_{x^2+y^2,inter}$ and $d_{x^2-y^2,inter}$ from NN inter-orbital AFM coupling. }\label{J2J1inter}
 \end{figure}

\begin{figure}%[t]
 \includegraphics[height=6.0cm]{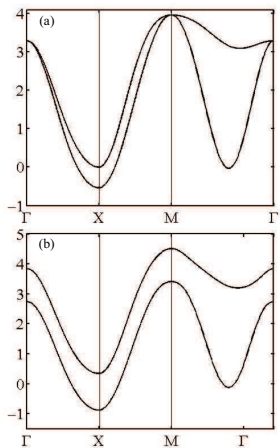}
 \caption{the band structure without SOC (a) and with SOC (b).} \label{bands}
 \end{figure}

The robustness of the extended $s$-wave is preserved even if  we include  other short range AFM couplings such as nearest-neighbor (NN) intra-orbital and inter-orbital ones or spin-orbital coupling (SOC). In the mean-field level, the NN intra-orbital AFM coupling favors $d_{x^2-y^2}$ pairing symmetry. As shown in Fig.~\ref{J1J2}, as soon as $J_2$ is larger than  the NN coupling $J_1$, the extended $s$-wave dominates.
The NN inter-orbital AFM coupling can induce four types of pairing symmetry, inter-orbital $s_{x^2+y^2}$, inter-orbital $d_{x^2-y^2}$ and inter-orbital $sinkx \pm i\cdot sinky$. Also, in the mean-field level, only inter-orbital $d_{x^2-y^2}$ pairing symmetry is favored by NN inter-orbital coupling and the extended $s$-wave induce by $J_2$ dominates while $J_2$ is larger than  the NN inter-orbital AFM coupling $J_{1,inter}$, as shown in Fig.\ref{J2J1inter}. The time reversal symmetry breaking inter-orbital $sinkx \pm i\cdot sinky$ does not exist in mean-field level.
Furthermore, the extended $s$-wave is also very robust against spin-orbital coupling.  Because of the absence of $p_z$ orbital contribution near Fermi surfaces, the onsite SOC Hamiltonian is given by $H_{SO}=-i\lambda/2\sum_{\mathbf{k}\sigma}\sigma (C^\dag_{X\textbf{k}\sigma}C_{Y\textbf{k}\sigma}-C^\dag_{Y\textbf{k}\sigma}C_{X\textbf{k}\sigma})$.
By fitting the first principle band structure with SOC with the two-band model, we get $\lambda\approx 1.1$ eV.  However, even if this SOC appears to be strong, it only lifts the degeneracy at X and $\Gamma$ points and makes no significant adjustment on Fermi surfaces.  We find that the extended $s$-wave is hardly affected. An only visible effect is that the SOC slightly increase the optimal doping level when we repeat the above mean-field calculation in the presence of SOC. Fig.\ref{bands} shows the band structure with and without SOC.

\begin{figure}%[b]
 \includegraphics[height=5.2cm]{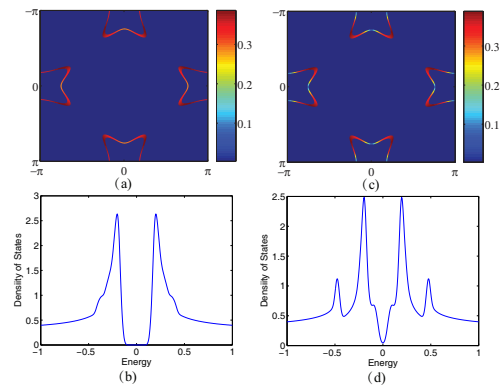}
 \caption{The SC gap distribution over the whole Fermi surface with $\Delta_X=0.2$ at $0.25$ electron doping and its corresponding DOS: (a),(b) for $s$-wave; (c),(d) for $d$-wave. }\label{overlap025}
 \end{figure}

\begin{figure}%[t]
 \includegraphics[height=5.2cm]{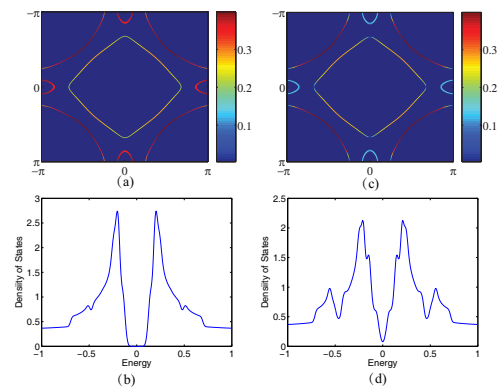}
 \caption{The SC gap distribution over the whole Fermi surface with $\Delta_X=0.2$ at $0.69$ electron doping and its corresponding DOS: (a),(b) for $s$-wave; (c),(d) for $d$-wave.}\label{overlap069}
 \end{figure}

In order to differentiate the extended $s$-wave from the $d$-wave or a conventional $s$-wave, we plot gap values around entire Fermi surfaces at different doping levels that are characterized by different topologies as shown in Figs.~\ref{overlap051}, \ref{overlap025} and \ref{overlap069}. For the extended $s$-wave, while the SC gap is developed throughout the Fermi surfaces, the gap has significant variations. The variations are different in three different Fermi surface topologies. In particular, at doping levels 0.51 and 0.69, the gap value around the corners of the square-shape Fermi surface at the center of Brillouin zone is significantly smaller than  the values on the other Fermi surfaces. This feature can allow us to distinguish the extended $s$-wave from the $d$-wave and a conventional $s$-wave. In the $d$-wave, as plotting in the same corresponding figures, there are gapless nodes at the above corners. In a conventional $s$-wave, the gap values should be more or less isotropic around Fermi surfaces. Experimental measurements by the standard scanning tunneling microscope (STM) or angle-resolved photoemission spectroscopy (ARPES) can provide a verdict.


\section*{Summary}
If the pairing in BiS$_2$-based superconductor is induced by the moderate electron-electron correlation, we show that similar to iron-based superconductors, the pairing symmetry is an extended $s$-wave, which is robust against doping and other parameter changes because of the quasi-one dimensional electronic structure of the materials. As the extended $s$-wave state is fully gapped with a significant gap variations on different parts of Fermi surfaces, our prediction can be directly justified or falsified by future experimental measurements.

We thank X. T. Zhang for extremely useful discussion. The work is
supported by "973" program (Grant No. 2012CV821400), as well as
national science foundation of China ( Grant No. NSFC-1190024). WFT is partially supported by NSC in Taiwan under grant no. 102-2112-M-110-009.

\end{document}